\newcommand{\la}{\langle}
\newcommand{\ra}{\rangle}

\newcommand{\beq}{\begin{eqnarray}}
\newcommand{\eeq}{\end{eqnarray}}

\newcommand{\gcon}{\la \frac{\alpha_s}{\pi} {G_{ij}^2} \ra }

\renewcommand{\theequation}{\thesection.\arabic{equation}}

\newcommand{\btem}{\bibitem}

\documentstyle[12pt]{article}

\begin{document}

\vspace{5pt}
 \centerline{\Large{\bf  Hadronic screening masses and }}
\vspace{10pt}
\centerline{\Large{\bf the magnetic gluon condensate at high temperature}}

\vspace{1.0cm}
\centerline{M. Ishii$^1$ and T. Hatsuda$^2$}

\vspace{0.5cm}

\noindent
$^1$ {\em Department of Physics,  Kyoto University, Kyoto 606-01, Japan}

\noindent
$^2$ {\em Institute of Physics, University
 of Tsukuba, Tsukuba, Ibaraki 305, Japan}

\vspace{2cm}

\centerline{Abstract}
 The hadronic
screening mass at high temperature ($T$) in QCD$_4$
 is examined on the basis of the
 QCD sum rules in (2+1) dimensions.
 Due to  the magnetic gluon condensate at high $T$
 which is expected to be nonvanishing,
 the screening mass deviates from  the asymptotic value $2\pi T$.
 Also, the screening mass in the vector (pseudo-vector) channel
 turns out to be heavier than that in the scalar (pseudo-scalar) channel.

\vskip10cm
e-mail:\ ishii@jpnyitp.bitnet,\ hatsuda@nucl.ph.tsukuba.ac.jp

\newpage

\setcounter{equation}{0}
\renewcommand{\theequation}{\arabic{equation}}

Quantum Chromodynamics (QCD) exhibits a transition from the
 hadronic phase to the quark-gluon plasma at finite temperature ($T$)
 and chemical potential.  In fact, numerical simulations on the lattice  show
a rapid growth of the internal energy near
 $ T_c \simeq 150$ MeV \cite{review}. Originally, it was expected that
   the system
 above $T_c$ behaves as a weakly interacting gas of
 quarks and gluons. However,
 there have arisen several indications against the simple
 picture at high $T$, which include
 the infrared breakdown of  perturbation theory
 \cite{linde},  area law behavior of the space-like Wilson loop \cite{borg},
  deviation  from the Stefan-Boltzmann law for pressure
 \cite{karsch},  non-vanishing gluon condensates  \cite{slee},
 and non-perturbative effects in  hadronic correlations
  \cite{hk85,kogut}.

  The purpose of this paper is to
  examine the relation between  two  non-perturbative effects
 at high $T$, i.e., the magnetic gluon-condensate and
  the hadronic screening masses.
   The latter is related to the
 meson masses in the QCD$_3$ + adjoint Higgs system which is a high $T$
 effective theory of QCD$_4$ (see \cite{pisarski}
 for the recent developments).
  We will use the
 QCD sum rules in (2+1) dimensions to study the
 spectra of the effective theory.
  Relation of our approach to the previous works \cite{shuryak,zahed}
 will be also discussed briefly.

 The hadronic screening mass $\mu(T)$ is defined by the space-like hadronic
correlation at finite $T$,
\beq
\label{correlation}
\Pi \equiv \la J(0,{\bf x}) J(0,{\bf y}) \ra  \sim
 e^{-\mu(T)|{\bf x}-{\bf y}|}
\ \ \ \ \  \ \  (|{\bf x}-{\bf y}| \rightarrow \infty ) \ \ ,
\eeq
where $\la \cdot \ra$ denotes thermal average and $J(\tau,{\bf x})$ is a
color-singlet hadronic operator.
In the following, we will focus on mesonic correlations
and take  $J=\bar{q}\gamma_0 q$ for the vector channel and
$J=\bar{q}q$ for the scalar
channel for simplicity.
\footnote{In the chirally symmetric phase above $T_c$,
 $\mu(T)$ takes the same  value in the same chiral
multiplet.  Also, we will not consider
the disconnected diagrams in (\ref{correlation}) for simplicity.
 The lattice QCD simulations currently available
also make this approximation \cite{kogut}.}
 If the  system is nearly a free quark-gluon gas at high $T$,
   $\mu (T) = 2 \pi T$ ($\pi T$ being the lowest Matsubara frequency for
 fermions) should be satisfied
in {\em both} channels  \cite{ioffe}.
  However, numerical simulations on the lattice show a sizable
 difference between $\mu(T)$ in the vector channel and that in the
 scalar channel, which indicates
 non-negligible correlation above $T_c$.
  This phenomenon could be explained by
 (i) perturbative gluon
exchange, in particular short-range spin-spin interaction
giving a  difference between vector channel and scalar channel, and/or
 (ii) non-perturbative gluonic contribution surviving even above $T_c$.
  We will focus our attention on (ii) in this paper and
  analyze the hadronic correlations under the non-perturbative
 background fields  using the  QCD sum rules.
 Since we will start with the high $T$ effective theory \cite{pisarski},
 our analysis will be applicable only at relatively high $T$.

 The fermionic part of the functional integral at finite $T$
under the  background gauge field $A_{\mu}$ reads
\beq
\label{partition}
Z({\bf A},A_4) = \int [dq d\bar{q}] \exp [- \int_0^{\beta} d\tau
 \int d^3x \ \  \bar{q} i \gamma_{\mu}^E D_{\mu}^E q ] \ \ ,
\eeq
where $\gamma$ matrices satisfy $\{\gamma_{\mu}^E,\gamma_{\nu}^E\} =
 -2 \delta_{\mu \nu} $ and the color indices are suppressed.
 All the non-perturbative effects are hidden in the gauge sector of the
 functional integral which is not shown here. At high $T$, the relevant
 gauge degree of freedom is the ``static" background $A_{\mu}(0,{\bf x})$
  \cite{pisarski}.
 Let us first
substitute the following Fourier decomposition into eq.(\ref{partition}):
$ q(\tau,{\bf x}) = T \sum_n
 e^{i\omega_n\tau} q(\omega_n,{\bf x})$ with
$\omega_n = 2\pi (n+ {1 \over 2})T$.
 By absorbing extra $\gamma_4^E$ into
the definition of $\bar{q}$ and rescale the quark fields as
$q \rightarrow \sqrt{T} q$, one immediately
 arrives at $Z$ written as a sum of
3-dimensional quarks with  ``quark masses'' $\omega_n$
 moving in the background field $A_{\mu} \equiv
({\bf A}({\bf x}),\phi({\bf x}))$:
\beq
\label{partition2}
Z({\bf A},\phi) = \int [dq d\bar{q}] \exp
 [-  \sum_n \bar{q}_n({\bf x})(i \Gamma_{i}^E D_i^E - \omega_n + g \phi)
 q_n({\bf x})]  \ \ ,
\eeq
where we have defined new $\gamma$ matrices $\Gamma_i^E
\equiv -i\gamma_0^E \gamma_i^E$
satisfying  $\{{\Gamma_i}^E,{\Gamma_j}^E\} =
-2 \delta_{ij} $.  For later convenience, we have not rescaled
 the background gauge fields, thus $g$ in eq.(\ref{partition2}) is the
 dimensionless coupling constant in QCD$_4$.
 Since $A_{\mu}$ is static ($\tau$ independent), there is no
transition among quarks with  different $n$.
 Thus, at high $T$, we can safely take $n=0$ sector to compute the
space-like correlation  (\ref{correlation}):
 Contributions from $n \neq 0$ sectors are exponentially
suppressed relative to the $n=0$ sector.
 In the following, we will omit the suffix $n$ assuming that
 $n=0$ sector dominates.

 By taking into account the redefinitions
 of the quark field we have made,
 eq.(\ref{correlation}) in the vector and scalar channel
 can be written as
\begin{eqnarray}
\label{correlation2}
\Pi_S & \propto  & \la \bar{q} i \gamma_4^E q ({\bf x})
    \bar{q} i \gamma_4^E q({\bf 0}) \ra
     \rightarrow e^{-m_{_S} |{\bf x}|} , \\
\label{correlation2*}
\Pi_V & \propto & \la \bar{q} i q ({\bf x}) \bar{q} i q({\bf 0}) \ra
   \ \ \ \ \ \ \    \rightarrow e^{-m_{_V} |{\bf x}|}.
\end{eqnarray}
 The vector (scalar) correlation in QCD$_4$ at finite $T$
is reduced to a scalar (skew scalar) correlation in the 3-dimensional
Euclidean space
at zero $T$. Our 4-component
             representation of the
            Dirac field in 3-dimensions
            contains
           two-layered 2-component spinors in 3-dimensions.
            This is why we have two kinds of scalars: $ \bar{q} q $ and
            $ \bar{q} \gamma_4^E q$. Similar 4-component
 spinor is adopted in \cite{jakiew} to study the
 chiral symmetry breaking in three dimensional QED.

 $\mu(T)$ in (\ref{correlation})  is identified with the masses
$m_{_{S,V}}$ defined in (\ref{correlation2},\ref{correlation2*}).
 A possible way to evaluate $m_{_{S,V}}$ is
to  make an analytic continuation of $\Pi_{S,V}$ into
(2+1)-dimensional Minkowski space and look for the lowest energy poles.
 This is analogous to the``funny space'' trick adopted
in ref.\cite{shuryak,zahed} where the temporal
 and one space direction is interchanged in
  QCD$_3$.

 An major assumption in our procedure
 is that the effective theory is in the confining phase and
  $\Pi_{S,V}$  has always isolated hadronic-poles in the
 Minkowski space. This assumption seems to be supported
 by the
  recent lattice QCD simulations where
  the effective theory at
high $T$ is shown to be in the confining phase and not in the
 Higgs phase, namely $\la \phi \ra =0$ \cite{reisz}.
 Thus, in the following,   we will set  $ \phi =0$ in
 (\ref{partition2}) and keep
 only  ${\bf A}(\bf x)$ which is essential for confinement in (2+1)-dimensions.

    To  extract the pole positions of $\Pi_{S,V}$ in the Minkowski space,
we are going to use QCD sum rules \cite{QSR} in (2+1)-dimensions which is
suitable to relate the spectral parameters and the background
fields (condensates).
 Here, $\la G_{ij}^2 \ra$ is the relevant operator
 with the lowest dimension and gauge invariance in our analysis.
 This is nothing but the magnetic condensate
$\la {\bf B}^2 \ra$ in 4 dimension.
 Lattice simulations of $\la {\bf B}^2 \ra$ in QCD$_4$ at $T> T_c$
show that it is non vanishing at least for  $T_c < T < (2 - 3) T_c$
\cite{slee}.

  Following the standard procedure of QCD sum rules, we carry out
the operator product expansion (OPE) of
 ${\rm Re}\Pi_{S,V}(q^2 \rightarrow - \infty)$ up to the lowest
non-trivial order as is shown in Fig.1. We
  have neglected the perturbative $\alpha_s$ corrections and took
 into account only the leading condensate $\la G_{ij}^2 \ra$.

\vspace{0.3cm}

\centerline{\fbox{Fig.1}}

\vspace{0.3cm}

 To improve the  OPE series, we apply the
 Borel transformation \cite{QSR} to
 ${\rm Re} \Pi_{V,S}(q^2)$, which results in
 $\tilde{\Pi}_{V,S}(M^2)$ with $M^2$ being the Borel mass:
\beq
\label{ope}
\tilde{\Pi}_S (M^2)
 & = & {3 \over 4\pi M^2} g_1(2 m) + \frac{\pi}{3} \gcon
                  \frac{M^4+16 m^2 M^2-16 m^2}{3 m M^6}
                   e^{-\frac{4 m^2}{M^2}} \\
\label{opex}
\tilde{\Pi}_V (M^2) & = & {-3 \over {4\pi M^2}}
                \{ g_1(2 m)-4 m^2 g_0(2 m) \}
                    + \frac{2 \pi}{3} \gcon
                  \frac{M^2+8 m^2}{3 m M^4}
                   e^{-\frac{4 m^2}{M^2}}
\eeq
where  $m \equiv \pi T$ is the lowest Matsubara
 frequency, and
\beq
g_0(x) = \int_{x^2}^{\infty} ds {1 \over \sqrt{s}} e^{-s/M^2}, \ \ \ \
g_1(x) = \int_{x^2}^{\infty} ds \sqrt{s} e^{-s/M^2} .
\eeq
As for the imaginary part of $\Pi_{S,V}$, we take
\beq
\label{ima}
{\rm Im} \Pi_S (s) &= & a_S \delta (s-{m_S}^2)
                  +\frac{3}{4} \sqrt{q^2} \ \theta (s-S_{0 S})\\
\label{imax}
{\rm Im} \Pi_V(s) &= &  a_V \delta (s-{m_V}^2)
                  -\frac{3}{4} {{q^2-4 m^2}\over \sqrt{q^2}}
                     \ \theta (s-S_{0 V}) ,
\eeq
where $a_S$ $(a_V)$ is a constant and $S_{0 S}$ $(S_{0 V})$ denotes
  the continuum threshold
  in the scalar (vector) channel. The structure of the
 continuum parts is  simply obtained from Fig.1(a)
  for large and positive $q^2$.

 Eq.(\ref{ope},\ref{opex}) and (\ref{ima},\ref{imax})
 satisfy the Borel transformed dispersion relation
\beq
\label{borels}
 \tilde{\Pi}(M^2)={1\over {\pi M^2}} \int {\rm Im}\Pi(s)e^{-\frac{s}{M^2}} ds,
\eeq
 The sum rule here are similar to that for the charm-quark system
in (3+1)-dimensions.
 The charm-quark mass corresponds to our ``Matsubara mass'' ($\pi T$) and
the 4-dimensional vacuum condensate $\la G_{\mu \nu}^2 \ra_0$ corresponds to
 our 3-dimensional condensate $\la G_{ij}^2 \ra$.

 From (\ref{borels}), one obtains
 the screening masses as
\begin{eqnarray}
\label{finaleq}
 (\frac{m_{_S}}{2 \pi T})^2 -1 & = &
    \frac{1}{p^2} \frac{ \bar{g}_{21} (p)-\bar{g}_{21}
                        (p \frac{\sqrt{S_{0 S}}}{2 m} )
                       -\frac{8\pi^2}{27} C\
                         p^5 (4-2 p^2) }
                        { \bar{g_1} (p)-\bar{g_1}
                        (p \frac{\sqrt{S_{0 S}}}{2 m} )
                       +\frac{8\pi^2}{27} C\
                         p^3 (1+4 p^2-p^4)} \\
\label{finaleqx}
  (\frac{m_{_V}}{2 \pi T})^2 -1 & = &
    \frac{1}{p^2} \frac{  \bar{g}_{21}(p)
                         - \bar{g}_{21} (p \frac{\sqrt{S_{0 V}}}{2 m})
                         -p^2 \{ \bar{g}_{10}(p)
                          - \bar{g}_{10} (p \frac{\sqrt{S_{0 V}}}{2 m}) \}
                       +\frac{16\pi^2}{27} 2 C
                          p^5 }
                       { \bar{g}_{10}(p) -
                           \bar{g}_{10} (p \frac{\sqrt{S_{0 V}}}{2 m})
                       -\frac{16\pi^2}{27} C
                         p^5 (1+2 p^2) }   ,
\end{eqnarray}
where $p \equiv 2 m/M = {2 \pi T}/ M$,
\beq
  \bar{g_{ij}} (p) = \bar{g_i} (p) -p^2 \bar{g_j} (p), \ \ \ \
  \bar{g_n} (p) =2 e^{p^2} \int_p^\infty x^{2n} e^{-x^2} dx ,
\eeq
and
\beq
\label{ratio}
C \equiv  \frac{\gcon}{(2m)^4} = \frac{\gcon}{(2 \pi T)^4} \ \ .
\eeq

 Since the magnetic gluon condensate still exists above $T_c$
 due to the non-perturbative effect in the space-like direction\cite{slee},
 we assume $C \neq 0$ and introduce a scaling factor $a$ relative to the
 vacuum magnetic-condensate in 4-dimensions,
\beq
\label{condensatev}
 \la \frac{\alpha_s}{\pi} {G_{ij}}^2 \ra
= {1 \over 2a}  \la \frac{\alpha_s}{\pi} {\bf B}^2 \ra_0 ,
\eeq
where $ \
\la \frac{\alpha_s}{\pi} {\bf B}^2 \ra_0 =
(1/2) \la \frac{\alpha_s}{\pi} G_{\mu \nu}^2 \ra_0
\simeq
 (1/2)(360\ {\rm MeV})^4 $ \cite{QSR}.
 Lattice QCD simulations suggests
  $a \simeq 1$ for $T_c < T < (2-3) T_c$
\cite{slee}, while the free quark-gluon gas implies $a = \infty$.

 Since the Borel mass is a fictitious
 parameter,   the l.h.s. of eq.(\ref{finaleq},\ref{finaleqx})
 should be insensitive to the change of  $p$ in a certain window
 $p_{min} < p < p_{max}$ (Borel window).
 We choose the window
 by the following conditions:
 (i) The first terms in the OPE (eqns.(\ref{ope},\ref{opex})) are
             more than three times larger than the second terms so that the
             expansions are good enough,
 and (ii)  The second terms in eqns.(\ref{ima},\ref{imax}) are more than three
             times smaller than the first terms so that final results are
             not sensitive to the detail structure of the continuum.
  The $S_0$'s in eqns.(\ref{ima},\ref{imax}) are chosen so as to make
$(m_{_{S,V}}/2 \pi T)^2$
least dependent on $p$   within this window.
 The minima of $m_{_{S,V}}$ as a function of $p$,
 which we always find in the windows, are taken as the physical
screening masses.
 Even if we adopt an average of $m_{S,V}(p)$ over the window
 as a physical screening mass,
 the results does not change much since the Borel curve is quite flat.

 The gluon condensate and $T^4$ always enter as a ratio in $C$ (see
 (\ref{ratio})). Therefore,
 as long as the good stability of $m_{_{S,V}}(p)$ is obtained
 as a function of $p$ by the suitable choice of $S_{0}$'s,
 an approximate scaling relation holds
\beq
m_{_{S,V}}(T,a) \simeq m_{_{S,V}}(Ta^{1/4},1).
\eeq
Thus we can cover all values of $a$ by rescaling $T$:
  The screening masses with
 large $a$ (small magnetic condensate)
 at fixed $T$ corresponds to that with
 $a=1$ (standard value suggested in lattice
 simulations) at larger $T$.

 The final results for $m_{_{S,V}}$ as a function of
 $Ta^{1/4}$ are shown in Fig. 2.
 $p$ dependence of $m_{_S}$ at $Ta^{1/4}=$0.4 GeV is also shown
 in Fig.3 to show the good stability of our result in the
 Borel window.

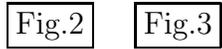
\vspace{0.3cm}

\centerline{\fbox{Fig.2} \ \ \  \fbox{Fig.3}}

\vspace{0.3cm}

 From Fig.2, the following final results follow:
\begin{description}
 \item[(1)] The screening masses in both channels deviate from
 the perturbative value  $2\pi T$ at moderate $T$ and approach asymptotically
 to $2 \pi T$ at large $T$. Furthermore,
  $m_{_V}-  m_{_S} \simeq 300\ {\rm MeV}$ for wide range of
 $T$.
 These effects are induced by the non-vanishing magnetic condensate
 above $T_c$.
 \item[(2)]  The screening mass in the scalar channel is
larger than $2 \pi T$ for the whole range of $T$.
 In fact, the coefficients of the gluon condensate $C$ in
  the r.h.s. of eq.(\ref{finaleq}) turn out to be positive
 in the Borel window, therefore the r.h.s. is always positive.
\end{description}

 The first feature (1), in particular, the
  positive gap $m_{_V} - m_{_S} = O(300\ {\rm MeV})$
 for wide range of $T$
  is  quantitatively consistent with the
 result of the lattice QCD simulations \cite{kogut}. Although
 the lattice data are limited to relatively low $T$
 region ($T < 2.5 T_c$) while our effective theory is valid
 at relatively high $T > (2-3) T_c$ \cite{pisarski},
 this agreement may give us a physical interpretation of the
 lattice data.
 The positive gap has been also
 obtained theoretically in ref.\cite{shuryak,zahed}
 where the spin-spin interaction in the ``funny'' space
 is essential.
 It is an open problem to understand the reason why
 our non-perturbative approach and their perturbative
  approach give similar result.

The result (2) is qualitatively different from the
 lattice data \cite{kogut} which show   $m_{_S} < 2 \pi T$.
 Within the approximation adopted in this paper, we
do not understand yet the origin of this difference.
 (Note that the funny space approach also shows
 $m_{_{S,V}} > 2 \pi T$  if the spin-spin interaction is neglected
 \cite{zahed}.)
 So far, we have neglected several effects such as
 the $\alpha_s$ corrections to Fig.1(a) from the static modes
 (${\bf A}$,$\phi$) as well as the
  non-static modes.
  Since the effective quark mass $m=\pi T$ is heavy
 at high $T$, the typical distance between $q$ and $\bar{q}$ is
 small. Thus the above corrections, in particular, the
 spin-independent attraction  could give rise to non-negligible
 reduction of the screening masses just like
  the perturbative color-Coulomb interaction in the
 heavy-quark systems such as $J/\psi$.
 QCD sum rules incorporating
  these effects are currently under investigation and will be reported
 elsewhere.

 In summary, we have studied effects of  the non-perturbative magnetic
 condensate at high $T$ on the hadronic correlations.
 The condensate gives rise to
  sizable corrections to the hadronic screening masses from their
  asymptotic value $2 \pi T$ in a channel dependent way.
 The positive gap $m_{_V} - m_{_S}$ and its magnitude are
  consistent with the current lattice data.

\vspace{3cm}

 The authors would like to thank S. H. Lee,  S. Huang, and
 J. Polonyi for useful discussions at the Institute for Nuclear Theory,
 Univ. of Washington. We also thank T. Kunihiro, T. Matsui and
 the members of the nuclear theory group at Kyoto Univ.
 for valuable comments. We are  particularly
 grateful for continuous encouragements by Prof. R. Tamagaki.

\newpage

\centerline{\bf{Figure captions}}

\vspace{1cm}

\noindent
Fig.1 OPE diagrams in QCD$_3$.  (a) perturbative contribution and
  (b) the power correction by the magnetic gluon condensate.

\vspace{0.8cm}

\noindent
 Fig.2  The screening masses devided by $2 \pi T$ against the scaled
 temperature $Ta^{1/4}$ (GeV).

\vspace{0.8cm}

\noindent
 Fig. 3 The behavior of $(m_{_{S}}/2\pi T)^2$ for the scalar
channel at
$Ta^{1/4}=0.4$ GeV in the window of $p$ defined in the text.

\end{document}